\documentstyle[prl,graphicx,aps]{revtex}
\begin{document}
\draft
\twocolumn[\hsize\textwidth\columnwidth\hsize\csname @twocolumnfalse\endcsname

\title{Deviation from Snell's Law for Beams Transmitted Near
the Critical Angle: Application to Microcavity Lasers}
\author{H.~E.~Tureci and A.~Douglas~Stone}

\address {Department of Applied Physics, P. O. Box 208284, Yale University,
New Haven, CT 06520-8284}
\date{hakan.tureci@yale.edu, douglas.stone@yale.edu}
\maketitle

\begin{abstract} We show that when a narrow beam
is incident upon a dielectric interface near the critical angle for
total internal reflection it will be transmitted into the far-field with an
angular deflection from the direction predicted by Snell's Law, due to a phenomenon
we call ``Fresnel Filtering''.  This effect can be quite large for the
parameter range relevant to dielectric microcavity lasers.
\end{abstract}
\pacs{{\bf OCIS codes: 140.3410,140.4780,230.3990,230.5750,260.2110,260.5740}}]


A promising approach to making high-Q optical micro-cavities and
micro-lasers is
to base them on totally-internally reflected modes of dielectric
microstructures.
This approach is currently under intense investigation with resonators and
lasers based on a range of shapes: disks, cylinders, spheres
\cite{changbook}, deformed
cylinders and spheres
\cite{science,gornik,chang,rex,nobel,arcscars} squares
\cite{poon} and hexagons  \cite{hex}.  Many different mode geometries
have been observed in
such resonators, e.g. whispering gallery modes \cite{changbook}, bow-tie
modes \cite{science,gornik}, triangle \cite{nobel,arcscars} and square
modes \cite{poon}.
Typically these modes correspond to ray trajectories which are
incident on the boundary of
the resonator above or at the critical angle in order to achieve
adequately high Q-values and
may correspond to periodic ray ``orbits'' (POs) which are either 
stable, unstable or
marginally stable. The natural and simplest method for predicting how 
such a mode will
emit or scatter light is simply to apply Snell's law to the underlying ray
orbit and follow the refracted ray into the far-field.  For a ray which is
incident at the critical angle this would imply emission in the direction
tangent to the emission point.  However in several recent experiments
very large deviations from this simple expectation were observed
\cite{science,arcscars}.
We show below that such observations may be explained as arising from
the angular
spread in the resonant mode around the PO, and the very
rapidly varying transmission probability as a function of angle near the
critical angle.  This ``filters'' out the angular components which are
totally internally reflected (TIR) and preferentially transmits those which are
far from TIR, leading to a net angular rotation of the outgoing radiation from the
tangent direction.  We call this effect Fresnel Filtering (FF).

The basic effect occurs for a bounded beam of arbitrary cross-section incident
from a semi-infinite medium of index $n$ into vacuum, although it will
be quantitatively altered in a resonator due to the curvature and/or
finite length
of the boundary.  We thus begin with the infinite planar example, which
we can solve analytically, before presenting numerical results for
quadrupolar asymmetric resonant cavities (ARCs)\cite{nature}.  There is a large
literature on reflection of a beam from a dielectric interface near
or above the
critical  angle, as the reflected beam exhibits the Goos-H\"{a}nchen
lateral shift as
well as other ``non-specular'' phenomena  \cite{GH2}.  However only
a few of these works address the transmission \cite{trans1,trans2} of the
beam and these tend to focus on the evanescent effects in the near
field; none appear to have identified the Fresnel Filtering effect and
its relevance to dielectric micro-cavity resonators.

\begin{figure}[t]
\centering
\includegraphics[width=\linewidth]{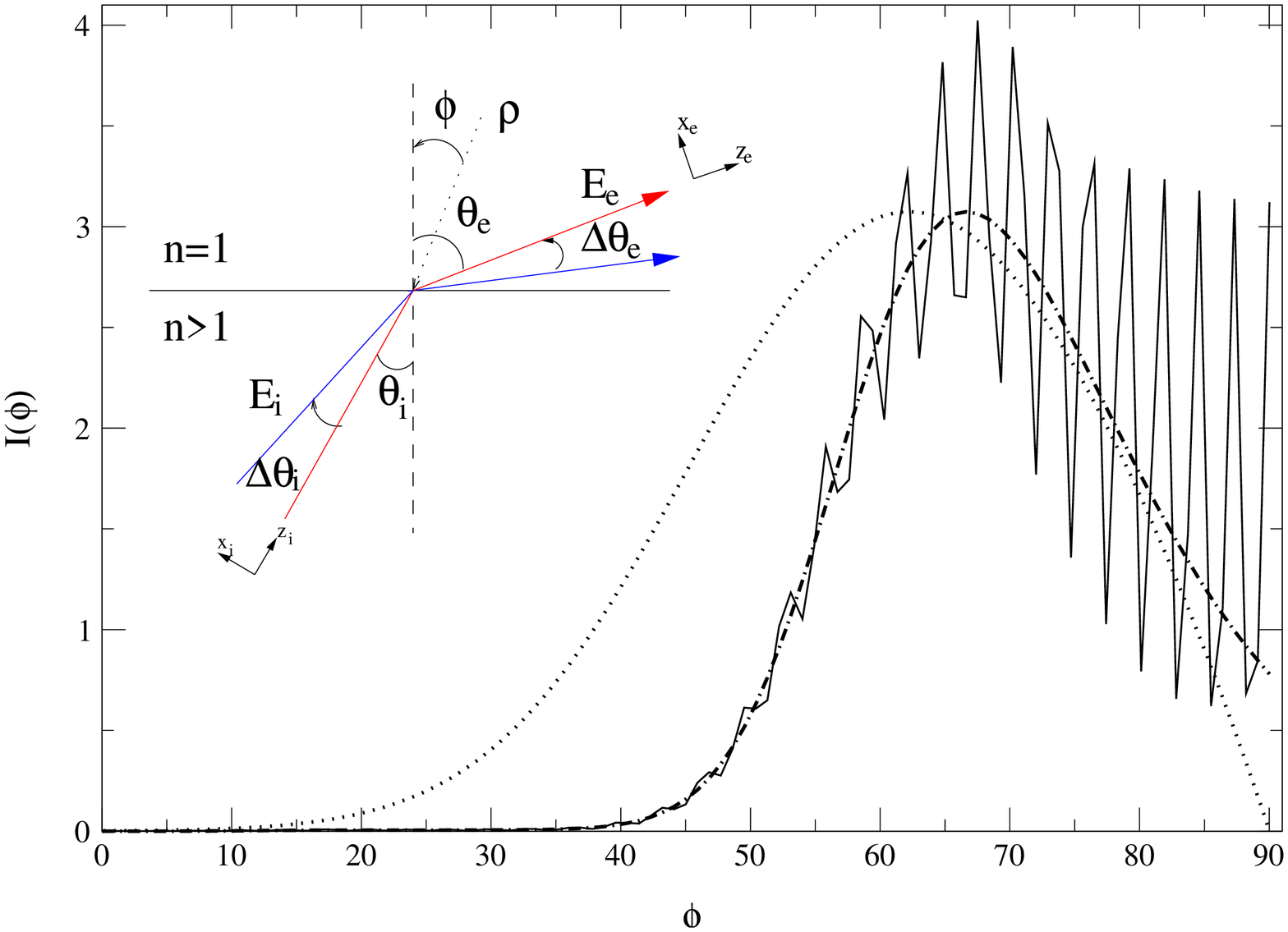}
\caption{Angular farfield intensity distributions 
$I(\phi)=|E_e(\phi)|^2$ for: (dotted line): critical
incidence on a planar interface with $n=1.56,\Delta = 8.82,z_0=5$ 
using the gaussian model (Eq. (\ref{eq:asymptote})). (solid line):  Exact quasi-normal mode with diamond geometry at $nk_or_o \approx  
90$, for a quadrupolar ARC with
$\varepsilon =0.1$, $n = 1.56$. (dot-dashed line): Chiral version of 
diamond resonance (see text) which
eliminates interference effects. Inset: Coordinates and variables for 
the GM calculation.}
\label{fig:profile}
\end{figure}

For simplicity, we consider a two-dimensional planar interface which separates
two semi-infinite regions with a relative dielectric
constant $n$.  Consider a beam $E_{i\alpha}$ incident from the denser
region with a central incidence angle $\theta_i$. We
will take the beam to be gaussian with a minimum beam waist $w$ (which we will use to scale all lengths henceforth) at a distance $z_o$
from the interface.  The basic effect is independent of the nature
of the input beam as long as it is focused and has significant angular
spread. The corresponding
Snell emission angle
$\theta_e$ (which is in general complex) is given by $n \sin \theta_i
= \sin \theta_e$.
${\cal S}_{i} : (x_{i},z_{i})$ and  ${\cal S}_{e} : (x_{e},z_{e})$
refer to coordinates
tied to the incident and refracted beams respectively (see
Fig.~\ref{fig:profile} inset). We will consider linearly polarized beams,
the corresponding beam fields $E_{\alpha}, \alpha = TM,TE,$ will then
denote the electric ($TM$) or the magnetic ($TE$)
fields normal to the plane of incidence.

Using the angular spectrum representation \cite{mandel},
the incident beam in ${\cal S}_{i}$ coordinates will consist of a
superposition of plane waves
of the same frequency $\omega$ with a gaussian distribution of
transverse wavevectors
$nk_{o}s$, where $s = \sin \Delta \theta_{i}$, $k_{o}=\omega / c_o$ is
the wavevector in vacuum and $\Delta \theta_{i}$ is the
deviation angle of the plane wave component from $\theta_{i}$:
\begin{equation}
E_{i\alpha}(x_{i},z_{i})=\frac{E_{o} \Delta}{2\sqrt{\pi}}
\int\limits_{- \infty}^{\infty}ds\,\exp\left[-
\left(\frac{\Delta}{2}\right)^{2} s^{2} + i \Delta\left(s x_{i} +
cz_i\right)\right]
\end{equation}
where $c=\sqrt{1-s^2}$ and the dimensionless width parameter $\Delta
= nk_{o}w$.


The beam on the $z>0$ side of the interface in polar coordinates
($\rho$, $\phi$) attached to the interface (after refraction) is
then given by the integral:
\begin{eqnarray}
E_{e\alpha}(\rho,\phi)& = &\frac{E_{o} \Delta}{2\sqrt{\pi}} \int
\limits_{- \infty}^{\infty}\,ds\,{\cal T}_{\alpha} (s) {\cal G}(s)\times\nonumber\\
&&\exp\left[i\frac{\Delta}{n}\rho\cos\left(\phi-\theta_{e}-\Delta\theta_{e}\right)\right]
\label{eq:ang}
\end{eqnarray}
Here $\Delta \theta_e$ is obtained from $n\sin (\theta_i+\Delta \theta_i)=\sin (\theta_e+\Delta \theta_e)$ and ${\cal G}(s)$ is given by:
\begin{equation}
{\cal G}(s) = \exp\left[- \left(\frac{\Delta}{2}\right)^{2} s^{2} + i
\Delta \sqrt{1-s^2}z_o\right]
\end{equation}
Evaluating this integral in the asymptotic farfield ($\rho \rightarrow
\infty$) using the saddle point method
we obtain our ``gaussian model'' (GM) for FF field:
\begin{equation}
E_{e\alpha}(\phi) = \frac{E_{o}
\Delta}{\sqrt{2i\frac{\Delta}{n}\rho}} \frac{\sqrt{1-s_{o}^2}\,
\cos\phi}{\sqrt{n^2-\sin^{2}\phi}} {\cal T}_{\alpha} (s_o) {\cal G}(s_o)
\exp\left( i\frac{\Delta}{n} \rho \right)
\label{eq:asymptote}
\end{equation}
where the transmission functions, evaluated at the relevant saddle
point 
\begin{eqnarray}
s_{o}(\phi)&=&\frac{1}{n}\left(\sin \phi \cos
\theta_{i} - \sin \theta_{i} \sqrt{n^{2} - sin^{2}\phi}\right)
\end{eqnarray}
are given by
\begin{eqnarray}
{\cal T}_{\alpha} [s_{o}(\phi)]& =& \frac{2 n
\sqrt{n^2-\sin^{2}\phi}}{ \mu \sqrt{n^2-\sin^{2}\phi} + n^{2}
\sqrt{1-\sin^{2}\phi}}
\end{eqnarray}
Here, $\mu = 1$ for $\alpha = TE$ and $\mu = n$ for $\alpha = TM$.
The relevant saddle point arises from setting to zero
the derivative of the cosine in the exponent of Eq.~ 
(\ref{eq:ang}); this saddle point value selects the angular 
component which refracts into the observation direction $\phi$ by 
Snell's law. However the amplitude factor obtained by gaussian
integration around the saddle point shifts the maximum of the
outgoing beam away from the Snell direction.  As noted, the effect occurs for
narrow beams with an arbitrary (non-gaussian) wavevector distribution 
${\cal B}(s)$; in
such a case the factor ${\cal G}(s_o)$ in Eq.~(\ref{eq:asymptote}) is 
replaced by
${\cal B}(s_0)$  (see e.g. ref. \cite{arcscars}).

Eq.~(\ref{eq:asymptote}) gives the angular beam profile in the
far-field, which is non-zero for any incident
angle
$\theta_i$, even $\theta_i > \theta_c = \sin^{-1}(1/n)$.  {\em
The key point is that the angular
maximum of this outgoing beam, $\phi_{max}$,
is in general not at the angle $\theta_e$ predicted by applying
Snell's law to the central incident beam direction $\theta_i$ }.
Instead, due to the unequal
transmission of the different angular components, the beam direction
is shifted by an angle $\Delta \theta_{FF}$ corresponding to
less refraction than expected from Snell's law.
This angular deflection can be quite large for incidence near $\theta_c$ in typical microcavity
resonators; in Fig.~\ref{fig:profile} the dashed line is the result 
of Eq.~(\ref{eq:asymptote}) for critical incidence, for which the 
Snell angle is $\phi =90^{\circ}$, but $\phi_{max} = 62^{\circ}$ 
giving $\Delta \theta_{FF}^c =
28^{\circ}$.  The farfield peak-shift $\Delta \theta_{FF}$ depends on 
the beam width $\Delta$ and on $n$; analysis of the
stationary phase solution gives the result that at $\theta_i=\theta_c$
\begin{equation}
\Delta \theta_{FF}^c \approx (2/\tan \theta_c)^{1/2}\Delta^{-1/2}
\label{eq:critical}
\end{equation}
which predicts $\Delta \theta_{FF}^c \approx 30^{\circ}$ for the parameters of
Fig.~\ref{fig:profile}.

Two technical points are in order here: First, while deforming the
contour of integration in Eq.~(\ref{eq:ang})
to the steepest 
descent path, depending on how
$\theta_i$,  $\theta_c$ and $\phi$ are situated, one might intercept 
branch cuts
due to the branchpoints $s = \pm 1$ and $s = \pm
\sin(\theta_{i}-\theta_{c})$.  Such branchpoint contributions are
well-studied for the reflected beam
shifts\cite{felsen-book,brekho}, but the contribution is 
subdominant with respect to the first order asymptotic term
derived in Eq.~(\ref{eq:asymptote}); we neglect such terms
here.  Second, there is another saddle point
$\tilde{s}_{o}=cos(\theta_{i})$ which corresponds to angular
components with grazing incidence to the interface.  Because
the Fresnel transmission factor vanishes for such components
$\tilde{s_0}$ only contributes to the integral at order
${\cal O} (\rho ^{-3/2})$.  This contribution only becomes important
very near $\phi = \pi/2$,  where the dominant
contribution from $s_0$ also vanishes, and again we neglect such terms in
the current  work.

\begin{figure}[hb]
\centering
\includegraphics[width=\linewidth]{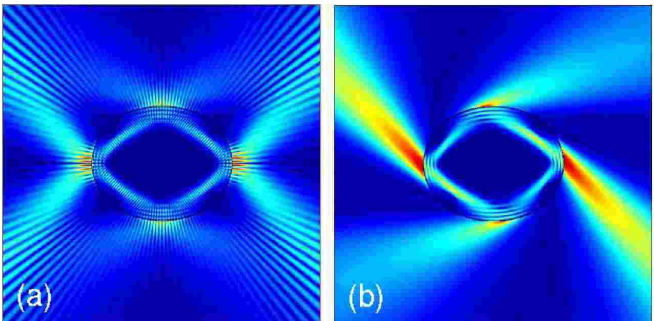}
\caption{(a) Field intensity plot (gray-scale) for a diamond 
resonance of the quadrupole
at critical incidence for the points at $\phi_w=0,\pi$, calculated
numerically at $nk_or_o \approx 90$, $n=1.56$, $\varepsilon = 0.1$. Note 
that there is negligible emission
from the upper and lower bounce points at $\phi_w=\pm 90^{\circ}$ 
because they are above the critical
angle (b) Chiral counterpart of this exact resonance, simulating a 
gaussian beam (see text).}
\label{fig:wf}
\end{figure}

Clearly the same Fresnel Filtering effect will occur in emission from
dielectric resonators with a magnitude similar to the planar case 
when the typical radius of curvature is much larger than $w$.  
As an example, we investigate the effect of FF on the farfield
emission pattern of quadrupolar ARCs
\cite{nature,nobel,science,arcscars}, dielectric cylinders
with cross-section given by $r(\phi_w) = r_{o}(1+\varepsilon\cos 
2\phi_w)$.  We study the exact
numerically generated quasibound TM modes of a resonator with $10 \%$
($\varepsilon = 0.1$) deformation for different
values of the refractive index $n$, focusing on resonances
based on the stable four-bounce (``diamond'') PO. The numerical method used is
a variant of the ``S-matrix method'' for billiards \cite{dietz,doron}.
If, as in this case, the relevant orbit is stable and we neglect
leakage, then it is possible to construct \cite{thebook}
approximate modes which are piecewise gaussian on each segment of the
PO.  From this theory one finds
that the effective beam waist in each segment will scale as
$\Delta = \xi \sqrt{nk_{o}r_o}$, where $\xi$ is a constant dependent
only on the stability matrix eigenvectors of that
particular segment, and $k_o$ is the quantized eigenvalue of the
mode. In Fig.~\ref{fig:wf} (a) we plot one representative
quasi-bound mode at
$n=1.56$; the corresponding far-field angular intensity 
is plotted in Fig.~\ref{fig:profile}.  

In both
figures one sees the rapid oscillations due to interference, but in
Fig.~\ref{fig:profile} one can see that the maximum of the intensity
is displaced from $\phi = 90^{\circ}$ as expected due to Fresnel
Filtering.  To compare the size of the effect to our
analytic theory for a planar interface it is convenient to eliminate the
interference by calculating the ``chiral" resonance
shown in  Fig.~\ref{fig:wf} (b).  This is the original resonance with
the negative angular momentum components
projected out, hence  mimicking a uni-directional beam.  When this is
plotted in Fig.~\ref{fig:profile} (dot-dashed line) it gives
the smooth envelope of the diamond resonance without the
oscillations.  We can regard this chiral resonance as a beam
incident at $\phi_w=0$ on the boundary and compare it to our planar
model.  The angle of incidence of the
``beam" with respect to the tangent plane is $\theta_i \approx
39^{\circ}$, and we have chosen $n$ so that $\theta_i=\theta_c$; hence naive ray
optics predicts tangent emission ($\phi_{max}=90^{\circ}$).  From 
Fig.~\ref{fig:profile} one
sees that the resonance emission is peaked at
$\phi_{max} \approx 66^{\circ}$, whereas the planar model gives a
similar envelope slightly shifted
with $\phi_{max} \approx 62^{\circ}$. In
evaluating the planar model we use $\Delta$ and $z_0$
as calculated from the gaussian theory of this diamond resonance,
hence we have no free parameters.  The
observed difference is likely due primarily to the effect of the
curvature of the boundary.

\begin{figure}[hb]
\centering
\vspace{6mm}
\includegraphics[width=\linewidth]{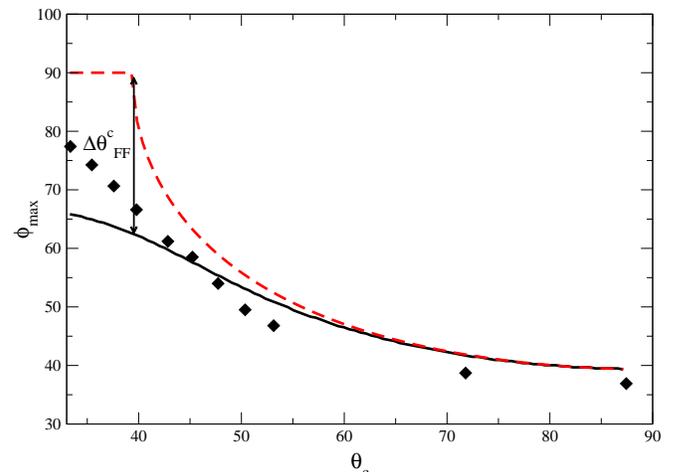}
\caption{Comparison of peak angular farfield values $\phi_{max}$ for 
varying critical angle $\theta_c
= \sin^{-1}(1/n)$. (Diamonds): exact resonances at $nk_or_o \approx 90$. (Solid line): GM calculation with $\Delta \approx 8.82$. (Dashed line):
Snell's law prediction: $\sin \phi_{max} = n\sin \theta_i$ where
$\theta_i \approx 39^{\circ}$. $\Delta \theta_{FF}^c$
designates the deviation from Snell's law at $\theta_c=\theta_i$.}
\label{fig:comparison}
\end{figure}

To evaluate systematically the Fresnel Filtering effect,
we have calculated the farfield peaks of the set of diamond resonances
while varying the index of refraction, so that the
critical angle is scanned through the PO incidence angle $\theta_i
\approx 39^{\circ}$. In order to remain as close as
possible to our GM with fixed $\Delta$ we have
chosen the resonances so that $nk_{o}r_{o}$ is approximately constant. In
Fig.~\ref{fig:comparison}, the exact numerical resonance peak is
compared to the calculated $\phi_{max}$ from Eq.~(\ref{eq:asymptote}) and
to the direction predicted by Snell's law. Clearly the deviation from 
Snell's law, $\Delta \theta_{FF}$,
varies with distance from $\theta_c$; further studies find that this region of
significant deviation decreases with increasing $\Delta$ as expected.

In conclusion,  we have shown that the transmission direction of a narrow
beam through a plane dielectric
interface can be quite different from the direction predicted by applying
Snell's law to the incident beam direction. This effect is due to a phenomenon
we call Fresnel Filtering and is of great importance in predicting the emission
patterns from resonances based
on periodic ray orbits in micro-cavity lasers.  This is true {\it even
when the size of the resonator $r_0$ is much larger than the wavelength}
and one might have expected ray optics to be quite good. 
Specifically the effective
beam waist for stable resonances scales as $\Delta
\propto \sqrt{nk_or_o}$, so from Eq. (\ref{eq:critical})
the deviation angle at critical incidence $\theta^c_{FF} \propto (nk_or_o)^{-1/4}$, and hence may be large for $nk_0r_0 \sim 10^2-10^3$
as in recent experiments on semiconductor ARC lasers
\cite{science,gornik,arcscars}.  

We acknowledge helpful discussions with H.~Schwefel, N.~Rex and R.~K.~Chang.  This work was supported by NSF grant DMR-0084501.

\end{document}